# Gamma radiation measurements and dose rates in commercially-used natural tiling rocks (granites)

**Michalis Tzortzis and Haralabos Tsertos***
*Department of Physics, University of Cyprus, Nicosia, Cyprus*

**Stelios Christofides and George Christodoulides**
*Medical Physics Department, Nicosia General Hospital, Nicosia, Cyprus*

## Abstract

The gamma radiation in samples of a variety of natural tiling rocks (granites) imported in Cyprus for use in the building industry was measured, employing high-resolution γ–ray spectroscopy. The rock samples were pulverized, sealed in 1-*litre* plastic Marinelli beakers, and measured in the laboratory with an accumulating time between 10 and 14 *hours* each. From the measured γ–ray spectra, activity concentrations were determined for $^{232}$Th (range from 1 to 906 $Bq\ kg^{-1}$), $^{238}$U (from 1 to 588 $Bq\ kg^{-1}$) and $^{40}$K (from 50 to 1606 $Bq\ kg^{-1}$). The total absorbed dose rates in air calculated from the concentrations of the three radionuclides ranged from 7 to 1209 $nGy\ h^{-1}$ for full utilization of the materials, from 4 to 605 $nGy\ h^{-1}$ for half utilization and from 2 to 302 $nGy\ h^{-1}$ for one quarter utilization. The total effective dose rates per person indoors were determined to be between 0.02 and 2.97 $mSv\ y^{-1}$ for half utilization of the materials. Applying dose criteria recently recommended by the EU for superficial materials, 25 of the samples meet the exemption dose limit of 0.3 $mSv\ y^{-1}$, two of them meet the upper dose limit of 1 $mSv\ y^{-1}$ and only one exceeds clearly this limit.

* *Corresponding author. E-mail address: tsertos@ucy.ac.cy, Fax: +357-22339060.*
*Department of Physics, University of Cyprus, P. O. Box 20537, 1678 Nicosia, Cyprus.*

## 1. Introduction

Gamma radiation from radionuclides which are characterized by *half-lives* comparable to the age of the earth, such as $^{40}K$ and the radionuclides from the $^{238}U$ and $^{232}Th$ series, and their decay products, represents the main external source of irradiation to the human body. The absorbed dose rate in air from cosmic radiation outdoors at sea level is about 30 $nGy\ h^{-1}$ (UNSCEAR 2000 Report). External exposures to gamma radiation outdoors arise from terrestrial radionuclides occurring at trace levels in all ground formations. Therefore, the natural environmental radiation mainly depends on geological and geographical conditions (Florou and Kritidis, 1992). Higher radiation levels are associated with igneous rocks, such as granite, and lower levels with sedimentary rocks. There are exceptions, however, as some shales and phosphate rocks have relatively high content of radionuclides (UNSCEAR 2000 Report).

*Granites* are the most abundant plutonic rocks of mountain belts and continental shield areas. They occur in great batholiths that may occupy thousands of square kilometres and are usually closely associated with quartz monzonite, granodiorite, diorite, and gabbro. They are extremely durable and scratch resistant; their hardness lends themselves for the stone to be mechanically polished to a high gloss finish. Their variety of colour and unique heat and scratch resistant properties makes them ideal for use as work-surface or as flooring or external and internal cladding. They mainly consist of coarse grains of quartz, potassium feldspar and sodium feldspar. Other common minerals that granites consist of, include mica and hornblende. Typical granites are chemically composed by 75 % silica, 12 % aluminium, less than 5 % potassium oxide, less than 5 % soda, as well as by lime, iron, magnesia, and titania in smaller quantities. Originally, it was widely believed that granites were formed mainly from magmatic differentiation of basaltic magma, evidence that was considered to indicate a metamorphic origin. However, because of the large quantities of granites that occur in nature, geologists believe now that most of the granites have been formed either by melting, partial melting, or metamorphism of deeply buried shale and sandstone. Granites, therefore, are the result of rapidly injected coalescing sheets of magma, each of which cooled independent of the other sheets. Evidence of intrusion or great mobility indicates an igneous origin that stems from melting of sediments, and consequently granite dykes are clearly igneous (Snelling and Woodmorappe, 1998).

In terms of natural radioactivity, granites exhibit an enhanced elemental concentration of uranium (U) and thorium (Th) compared to the very low abundance of these elements observed in the mantle and the crust of the Earth. Geologists provide an explanation of this behaviour in the course of partial melting and fractional crystallisation of magma, which enables U and Th to be concentrated in the liquid phase and become incorporated into the more silica-rich products. For that reason, igneous rocks of granitic composition are strongly enriched in U and Th (on an average 5 *ppm*[†] of U and 15 *ppm* of Th), compared to rocks of basaltic or ultramafic composition (< 1 *ppm* of U) (Faure, 1986; Mınager et al., 1993).

---

[†] *parts per million*

In this paper, the results from gamma radiation measurements in samples of a variety of natural tiling rocks imported in Cyprus and sold under the commercial name of “granites” are presented. These results are of general interest since such rocks are globally used as building and ornamental materials. The measurements have been carried out in the Nuclear Physics Laboratory of the Department of Physics, University of Cyprus, using a high-resolution $\gamma$–ray spectroscopic system.

## 2. Experimental method

### 2.1 Gamma-ray detection system

A stand-alone high-resolution spectroscopic system is used for the measurement of the energy spectrum of the emitted gamma rays in the energy range between 50 *keV* and 3000 *keV*. The system consists of a high-purity germanium (HPGe) detector (coaxial cylinder of 55 *mm* in diameter and 73 *mm* in length) with an efficiency of 30 %, relative to a 3″×3″ NaI(T*l*) scintillator. The spectroscopic system is linked with a Multi-Channel Buffer (MCB) which is a PC-based plug-in PCI card consisting of an 8k Analogue-to-Digital Converter (ADC). An advanced Multi-Channel Analyser (MCA) emulation software (MAESTRO-32) enables data acquisition, storage, display and online analysis of the acquired gamma spectra.

The detector is surrounded by a graded-Z cylindrical shield consisting of lead, iron, and aluminium with thickness of 5 *cm*, 1 *cm*, and 1 *cm*, respectively, which provides an efficient suppression of the background gamma radiation present at the laboratory site. The energy-dependent detection photopeak efficiency has been determined using a calibrated $^{152}$Eu gamma reference source[‡], sealed in a standard Marinelli beaker with 85

[‡] Manufactured by AEA Technology QSA GmbH, Germany and calibrated by the German calibration laboratory for measurements of radioactivity (DKD).

*mm* bore diameter, an active volume of 1000 *ml*, and an average density of 1 $g\ cm^{-3}$. It is characterised by initial activity of 10.1 $kBq\ kg^{-1}$. Considering an uncertainty of 3% in the source activity, the mean uncertainty in the calculated efficiency is estimated to be about 5%.

The energy resolution (FWHM) achieved in the measurements is 1.8 *keV* at the 1.33 *MeV* reference transition of $^{60}$Co. Depending on the peak background and the counting time of each measured spectrum, the calculated Minimum Detectable Activity (MDA) varies within the range of 0.02 - 0.2 $Bq\ kg^{-1}$ for both $^{232}$Th and $^{238}$U, and within the range of 0.05 - 0.5 $Bq\ kg^{-1}$ for $^{40}$K.

## 2.2 Sample preparation and counting

A total of 28 different kinds of “granites” of those imported in Cyprus have been collected (Table 1). The measured samples were pulverized, sieved through 0.2 *mm* mesh, sealed in standard 1000 *ml* Marinelli beakers, dry-weighed and stored for four weeks before counting in order to allow the reaching of equilibrium between $^{226}$Ra and $^{222}$Rn and its decay products. However, significant non-equilibrium is uncommon in rocks older than $10^6$ *years*, and the $^{232}$Th series may be considered in equilibrium in most geological environments (Chiozzi et al., 2002). Each sample was put into the shielded HPGe detector and measured for an accumulating time between 10 and 14 *hours*. A typical spectrum is shown in Fig. 1. Prior to the samples measurement, the environmental gamma background at the laboratory site has been determined with an empty Marinelli beaker under identical measurement conditions. It has been later subtracted from the measured γ–ray spectra of each sample.

The offline analysis of each measured γ–ray spectrum has been carried out by a dedicated software programme (GammaVision-32), which performs a simultaneous fit to all the statistically significant photopeaks appearing in the spectrum. Menu-driven reports were generated which included the centroid channel, energy, net area counts, background counts, intensity and width of identified and unidentified peaks in the spectrum, as well as peak and average activity in *Bq kg*$^{-1}$ for each detected radionuclide. The results of the reported $^{232}$Th, $^{238}$U and $^{40}$K activity concentrations obtained for each of the measured samples together with their corresponding total uncertainties are summarised in Table 1. It is noted here that no other radionuclide than naturally occurring were detected in the measured samples and that the small contribution of the environmental γ–ray background at the laboratory site has been subtracted from the spectra of the measured samples.

## 2.3 Calculation of activity concentrations

Calculations of count rates for each detected photopeak and radiological concentrations (activity per mass unit or specific activity) of detected radionuclides depend on the establishment of secular equilibrium in the samples. Since secular equilibrium was reached between $^{232}$Th and $^{238}$U and their decay products, the $^{232}$Th concentration was determined from the average concentrations of $^{212}$Pb and $^{228}$Ac in the samples, and that of $^{238}$U was determined from the average concentrations of the $^{214}$Pb and $^{214}$Bi decay products (Hamby and Tynybekov, 2000). Thus, an accurate radionuclide concentration of $^{232}$Th and $^{238}$U was determined, whereas a true measurement of concentration was made. The specific activity (in *Bq kg*$^{-1}$), $A_{Ei}$, of a nuclide *i* and for a peak at energy *E*, is given by:

$$A_{Ei} = \frac{N_{Ei}}{\varepsilon_E \times t \times \gamma_d \times M_s} \qquad (1)$$

where $N_{Ei}$ is the Net Peak Area of a peak at energy *E*, $\varepsilon_E$ the detection efficiency at energy *E*, *t* the counting live-time, $\gamma_d$ the gamma ray yield per disintegration of the specific nuclide for a transition at energy *E*, and $M_s$ the mass in *kg* of the measured sample. If there is more than one peak in the energy range of analysis, then the peak activities are averaged and the result is the weighted average nuclide activity.

The total uncertainty ($\sigma_{tot}$) of the calculated activity values (Table1) is composed of the counting statistical ($\sigma_{st}$) and weighted systematic errors ($\sigma_{sys,i}$) calculated by the following formula (EG&G ORTEC, 1999):

$$\sigma_{tot} = \sqrt{\sigma_{st}^2 + 1/3\sum_i \sigma_{sys,i}^2} \qquad (2)$$

The systematic uncertainties considered include: the uncertainty of the source activity (3%), the uncertainty in the efficiency fitting function (1-10%), and uncertainties in the nuclide master library used (1-2%). As can be seen in Table 1, the total uncertainty in the calculated activity concentration is in most cases below 5%.

### 2.4 Derivation of the absorbed dose rates and effective dose rates

If naturally occurring radioactive nuclides are uniformly distributed, dose rates, D, in units of *nGy* $h^{-1}$ can be calculated by the following formula (Kohshi et al., 2001):

$$D = A_{Ei} \times C_F \qquad (3)$$

where $A_{Ei}$ is the activity concentration measured in *Bq* $kg^{-1}$, and $C_F$ is the dose conversion factor (absorbed dose rate in air per unit activity per unit of soil mass, in units of *nGy* $h^{-1}$ *per Bq* $kg^{-1}$).

Dose conversion factors have been extensively calculated during the last forty years by many researchers. In the present work, the considered dose rate conversion factors for the $^{232}$Th and $^{238}$U series, and for $^{40}$K, used in all dose rate calculations are those determined by Saito et al. (1990) which have been used extensively for all similar calculations in the UNSCEAR 1993 Report. It should be pointed out here that, using this calculation, the dose rate for the $^{232}$Th and $^{238}$U series is the average of the respective radiological concentrations multiplied by the conversion factors corresponding to each series. The total dose rate for each of the measured samples is the sum of the dose rates contributed by both series of $^{232}$Th and $^{238}$U, and by $^{40}$K.

The building materials act as sources of radiation and also as shields against outdoor radiation (UNSCEAR 1993 Report). In massive houses made of different building materials such as stone, bricks, concrete or granite, the factor that mainly affects the indoor absorbed dose is the activity concentrations of natural radionuclides in those materials, while radiation emitted by sources outdoors is efficiently absorbed by the walls. Consequently, dose rates in air indoors will be elevated accordingly to the concentrations of naturally occurring radionuclides used in construction materials. In order to facilitate the calculation of dose rates in air from different combinations of the three radionuclides in building materials and by applying the appropriate conversion factors, an activity utilization index is constructed that is given by the following expression:

$$\left(\frac{C_{Th}}{A_{Th}} f_{Th} + \frac{C_U}{A_U} f_U + \frac{C_K}{A_K} f_K\right) w_m \qquad (4)$$

where $C_{Th}$, $C_U$ and $C_K$ are actual values of the activities per unit mass ($Bq\ kg^{-1}$) of $^{232}$Th, $^{238}$U, and $^{40}$K in the building materials considered; $f_{Th}$, $f_U$ and $f_K$ are the fractional contributions to the total dose rate in air due to gamma radiation from the actual concentrations of these radionuclides. In the NEA 1979 Report, typical activities per unit mass of $^{232}Th$, $^{238}U$, and $^{40}K$ in building materials $A_{Th}$, $A_U$ and $A_K$ are referred to be 50, 50 and 500 $Bq\ kg^{-1}$, respectively.

The activity utilization index is, finally, weighted for the mass proportion of the building materials in a house by being multiplied by a factor $w_m$ that represents the fractional usage of those materials in the dwelling with the characteristic activity. To be more specific, full mass utilization ($w_m$=1) of a given material implies that all building materials used in a model masonry house are composed of this specific material. Half mass utilization ($w_m$=0.5) means that 50% of the masonry mass is composed of the material considered, and so on. For full mass utilization of a model masonry house ($C_{Th} = C_U = 50\ Bq\ kg^{-1}$ and $C_K = 500\ Bq\ kg^{-1}$), the activity utilization index is unity by definition and is deemed to imply a dose rate of 80 $nGy\ h^{-1}$ (UNSCEAR 1993 Report).

Finally, in order to estimate the annual effective doses, one has to take into account the conversion coefficient from absorbed dose in air to effective dose and the indoor occupancy factor. In the UNSCEAR recent reports (1988, 1993, 2000), a value of 0.7 $Sv\ y^{-1}$ was used for the conversion coefficient from absorbed dose in air to effective dose received by adults, and 0.8 for the indoor occupancy factor, implying that 20 % of time is spent outdoors, on average, around world. The effective dose rate indoors, $H_E$, in units of *mSv per year,* is calculated by the following formula:

$$H_E = D \times T \times F \qquad (5)$$

where D is the calculated dose rate (in $nGy\ h^{-1}$), T is the indoor occupancy time ($0.8 \times 24\ h \times 365.25\ d \cong 7010\ h\ y^{-1}$), and F is the conversion factor ($0.7\times10^{-6}\ Sv\ Gy^{-1}$).

In Table 2, the results obtained for the activity utilization index and the total absorbed dose rate in air due to gamma radiation for various fractional masses of the 28 “granite” samples, as well as the indoor effective dose assessment for a specific fractional mass for each sample are presented.

## 3. Results and discussion

Activity concentrations of $^{232}$Th ranged from 1 to 906 *Bq kg*$^{-1}$, of $^{238}$U from 1 to 588 *Bq kg*$^{-1}$ and of $^{40}$K from 50 to 1606 *Bq kg*$^{-1}$. From the 28 samples measured in this study, “Café Brown” appears to present the highest concentrations for all the elements investigated, reaching levels of 906 *Bq kg*$^{-1}$ for $^{232}$Th, 588 *Bq kg*$^{-1}$ for $^{238}$U, and 1606 *Bq kg*$^{-1}$ for $^{40}$K. “Rosso Balmoral” and “New Imperial” exhibit the second and third highest concentration of $^{232}$Th reaching 490 *Bq kg*$^{-1}$ and 273 *Bq kg*$^{-1}$, respectively, and of $^{238}$U reaching 162 *Bq kg*$^{-1}$ and 285 *Bq kg*$^{-1}$, respectively, while “Upatuba” appears to present the second higher concentration of $^{40}$K reaching 1581 *Bq kg*$^{-1}$. All measured samples except of two, named “Astudo” and “Nero Africa”, show concentrations of $^{40}$K above the value of 1000 *Bq kg*$^{-1}$. In addition, 13 samples appear to present concentration of $^{232}$Th higher than 100 *Bq kg*$^{-1}$, while only 6 samples exhibit concentration of $^{238}$U that surpass the above limit. “Nero Africa” demonstrates the lowest concentrations for all the investigated elements, showing nearly zero net activity levels for $^{232}$Th and $^{238}$U, and levels of 50 *Bq kg*$^{-1}$ for $^{40}$K. Analytical results for the activity concentrations of $^{232}$Th, $^{238}$U and of $^{40}$K determined for each of the measured samples together with their total uncertainties are presented in Table 1.

The measured activity concentrations of $^{232}$Th, $^{238}$U and $^{40}$K can be converted into total elemental concentrations of Th, U (in *ppm*) and of K (in *percent)*, respectively (Tzortzis et al., 2002). The extracted values for the elemental concentration are: for Th (range from 0.6 to 223.0 *ppm*), for U (from 0.1 to 47.7 *ppm*) and for K (from 0.2 to 5.3%). In all cases, the lower values correspond to sample No.15 ("Nero Africa") and the higher ones to sample No.25 ("Café Brown"). The results are summarised in Fig. 2.

In estimating the effect of using atypical materials such as granites or other natural tiling rocks, it is necessary to determine the fractional utilization by mass for each of the measured samples and identify the associated dose rate. The activity utilization index estimated using the fractional contribution to the dose rate from the three radionuclides for the 28 samples are presented in Table 2. As the activity concentration of the three radionuclides ($^{232}$Th, $^{238}$U, $^{40}$K) and their corresponding fractional contribution to the total dose rate vary from sample to sample, the activity utilization index ranges from 0.09 for sample "Nero Africa" to 15.11 for sample "Café Brown". It should be noted that, 22 of the measured samples exhibit an activity utilization index that ranges from 1.0 to 3.0, four of them surpass the above-mentioned range and only two samples show values under that range ($\leq 1.0$).

The total absorbed dose rates calculated from the concentrations of the nuclides of the $^{232}$Th and $^{238}$U series, and of $^{40}$K, for full utilization of the measured tiling materials range from 7 to 1209 $nGy\ h^{-1}$, values that are decreased according to the fractional utilization of the materials. In round terms and for full utilization, 23 of the samples exhibit dose rates that range from 100 to 400 $nGy\ h^{-1}$, two exhibit dose rates under the typical limit of 100 $nGy\ h^{-1}$ ("Nero Africa" with 7 $nGy\ h^{-1}$ and "Astudo" with 43 $nGy\ h^{-1}$) and only three of the samples exhibit values over the limit of 400 $nGy\ h^{-1}$ ("Café Brown" with 1209 $nGy\ h^{-1}$, "Rosso Balmoral" with 619 $nGy\ h^{-1}$ and "New Imperial" with 409 $nGy\ h^{-1}$). A summary of this analysis is presented in Fig. 3, where the frequency (in *percent*) of total absorbed dose rates for full utilization of the 28 measured samples is plotted.

The analytical results for the total absorbed dose rates in air for each of the measured samples and for various fractional masses indicated are also given in Table 2. For comparison, measurements in former Czechoslovakia, in houses with outside walls containing uraniferous coal slag, gave values approaching 1,000 $nGy\ h^{-1}$ (Thomas et al., 1993), while measurements in a granite region of the United Kingdom, where some of the houses are made of local stone, gave 100 $nGy\ h^{-1}$ (Wrixon et al., 1988).

The relative contribution to total absorbed dose due to $^{232}$Th ranges from 1 % for sample "Nero Africa" to 69 % for "Rosso Balmoral", due to $^{238}$U ranges from 4 % for "Multi-colour" to 30 % for "Café Brown" and due to $^{40}$K ranges from 7 % for "Café Brown" to 92 % for "Nero Africa". The fractional contribution to total absorbed dose of the three radionuclides, $^{232}$Th, $^{238}$U and $^{40}$K, is plotted in Fig. 4, for all the samples studied.

Finally, the annual effective doses indoors estimated, using equation (5) and 0.5 utilization of material, range from 0.02 $mSv\ y^{-1}$ for sample "Nero Africa" to 2.97 $mSv\ y^{-1}$ for "Café Brown". The latter is only one sample from all those examined that exceeds the average worldwide exposure of ~2.4 $mSv\ y^{-1}$ due to natural sources (UNSCEAR 2000 Report). It is clear that, for full utilization of the materials, the associated effective dose rates would be by a factor of two higher than those illustrated in Table 2 ($w_m$ = 0.5) and by a factor of two lower for one quarter utilization of the samples. However, the simple irradiation geometry and rounded values considered do not allow this approach to be more than a rough approximation.

According to most recent regulations and especially the recommendation No. 112 issued by the European Union in 1999 (EC, 1999), building materials should be exempted from all restrictions concerning their radioactivity, if the excessive gamma radiation due to those materials causes the increase of the annual effective dose received by an individual by a maximum value of 0.3 *mSv*. Effective doses exceeding the dose criterion of 1 $mSv\, y^{-1}$ should be taken into account in terms of radiation protection. It is therefore recommended that controls should be based on a dose range of 0.3–1 $mSv\, y^{-1}$, which is the building material gamma dose contribution to the dose received outdoors. In order to examine whether a building material meets these two dose criteria, the following gamma activity concentration index (*I*) is derived:

$$I = \frac{C_{Th}}{200\ \mathrm{B}q\, kg^{-1}} + \frac{C_U}{300\ \mathrm{B}q\, kg^{-1}} + \frac{C_K}{3000\ \mathrm{B}q\, kg^{-1}} \qquad (6)$$

where $C_{Th}$, $C_U$, $C_K$ are the Th, U and K activity concentrations ($Bq\, kg^{-1}$) in the building material, respectively. For superficial and other building materials with restricted fractional mass usage, such as those studied in this work, the exemption dose criterion (0.3 $mSv\, y^{-1}$) corresponds to an activity concentration index $I \leq 2$, while the dose criterion of 1 $mSv\, y^{-1}$ is met for $I \leq 6$ (EC, 1999). This approach has initially been developed by the Radiation Protection Authorities in the Nordic Countries (Nordic, 2000)[§] and is generally accepted by the EU member states and many other countries.

Based on the activity concentration index calculated according to equation (6), 25 of the samples exhibit $I \leq 2$, two of them, “Rosso Balmoral” with $I \cong 3.5$ and “New Imperial” with $I \cong 2.7$, show values $2 \leq I \leq 6$, and only one, “Café Brown” with $I \cong 7.5$, exceeds clearly the dose limit of 1 $mSv\, y^{-1}$. It should be noted that quite similar conclusions are drawn by considering the effective dose rates calculated according to equations (4, 5) and using a mass utilization factor of $w_m$ = 0.25, which seems to be appropriate for superficial materials.

[§] The reference to this report by one of our external reviewers is gratefully acknowledged.

## 4. Conclusions

Exploitation of high-resolution γ–ray spectroscopy provides a sensitive experimental tool in studying natural radioactivity and determining elemental concentrations and dose rates in various rock types. Most of the tiling rock "granite" samples studied in this work reveal high values for the activity and elemental concentrations of Th, U and K, thus contributing to high absorbed dose rates in air. In general, the extracted values are distinctly higher than the corresponding population-weighted (world-averaged) ones, and, in some cases (e.g. "Café Brown", "Rosso Balmoral", "New Imperial" etc.), they lie outside the typical range variability of reported values from world-wide areas due to terrestrial gamma radiation, given in the recent UNSCEAR 2000 Report.

In addition, according to the dose criteria recently recommended by the European Union (EC, 1999), 25 of the samples meet the exemption dose limit of 0.3 $mSv\ y^{-1}$, two of them "Rosso Balmoral" and "New Imperial" meet the upper dose limit of 1 $mSv\ y^{-1}$, and only one "Café Brown" exceeds clearly this limit. These statements are valid as long as the use of these materials is restricted (superficial materials).

## Acknowledgements

This work is conducted with financial support from the Cyprus Research Promotion Foundation (Grant No. 45/2001) and partially by the University of Cyprus. We would also like to thank the Cyprus Association of Manufacturers, which provided us with the original granite materials from which the samples are prepared and gave us information about their origin.

# TABLE CAPTIONS

**Table 1.** Activity concentrations of $^{232}$Th, $^{238}$U and $^{40}$K of natural tiling rocks (granites) imported and used in Cyprus.

| | Commercial name | Country of origin | Concentration ± Total Error (*Bq kg$^{-1}$*) | | |
|---|---|---|---|---|---|
| | | | $^{232}$Th | $^{238}$U | $^{40}$K |
| 1 | Bianco Perla | Italy | 37 ± 2 | 57 ± 2 | 1228 ± 48 |
| 2 | Santa Cecilia | Brazil | 85 ± 2 | 45 ± 1 | 1435 ± 55 |
| 3 | Blue Paradise | Brazil | 92 ± 3 | 15 ± 1 | 1246 ± 48 |
| 4 | Blue Pearl | Belgium | 77 ± 2 | 68 ± 2 | 1129 ± 44 |
| 5 | Verte Brazil | Brazil | 121 ± 3 | 5 ± 1 | 1200 ± 47 |
| 6 | Upatuba | Africa | 21 ± 1 | 17 ± 1 | 1581 ± 61 |
| 7 | Verte Eukaliptos | Brazil | 26 ± 1 | 45 ± 1 | 1522 ± 59 |
| 8 | Red Africa | Africa | 113 ± 2 | 57 ± 1 | 1360 ± 38 |
| 9 | Tropical Japorana | Brazil | 17 ± 1 | 13 ± 1 | 1048 ± 30 |
| 10 | Astudo | Africa | 32 ± 1 | 18 ± 1 | 254 ± 11 |
| 11 | Baltic Brown | Brazil | 136 ± 4 | 102 ± 3 | 1520 ± 58 |
| 12 | Rosso Balmoral | Holland | 490 ± 13 | 162 ± 5 | 1540 ± 60 |
| 13 | Rossa Porino | Italy | 172 ± 5 | 103 ± 3 | 1424 ± 55 |
| 14 | Giallo Penere | Brazil | 82 ± 2 | 31 ± 1 | 1230 ± 47 |
| 15 | Nero Africa | Africa | 0 ± 1 | 1 ± 1 | 50 ± 3 |
| 16 | Rosa Beta | Italy | 69 ± 2 | 40 ± 1 | 1124 ± 44 |
| 17 | White Arabesco | N/A | 146 ± 4 | 108 ± 3 | 1359 ± 52 |
| 18 | Saint Tropez | Brazil | 40 ± 1 | 8 ± 1 | 1021 ± 40 |
| 19 | Kinawa | Brazil | 101 ± 3 | 58 ± 2 | 1168 ± 45 |
| 20 | Multi-colour | N/A | 82 ± 2 | 10 ± 1 | 1486 ± 57 |
| 21 | Capao Bonito | Brazil | 190 ± 5 | 84 ± 2 | 1313 ± 51 |
| 22 | New Imperial | N/A | 273 ± 7 | 285 ± 8 | 1273 ± 49 |
| 23 | Juparana | Brazil | 265 ± 7 | 35 ± 1 | 1446 ± 56 |
| 24 | Grand Paradisso | N/A | 51 ± 1 | 29 ± 1 | 1013 ± 39 |
| 25 | Café Brown | Brazil | 906 ± 24 | 588 ± 16 | 1606 ± 62 |
| 26 | Rosa Ghiandone | Italy | 89 ± 3 | 57 ± 2 | 1047 ± 41 |
| 27 | Jacaranda | Brazil | 147 ± 4 | 68 ± 2 | 1031 ± 40 |
| 28 | Colibri | Brazil | 155 ± 4 | 53 ± 1 | 1365 ± 53 |

**Table 2.** Gamma-ray exposure in natural tiling rocks (granites) imported and used in Cyprus.

| Sample number | Activity utilization index | Absorbed dose rate** in air for indicated fractional mass of building material ($nGy\ h^{-1}$) | | | | Effective dose rate indoors ($mSv\ y^{-1}$) $w_m$= 0.5 |
|---|---|---|---|---|---|---|
| | | $w_m$ = 1.0 | 0.75 | 0.5 | 0.25 | |
| 1 | 1.72 | 137 | 103 | 69 | 34 | 0.34 |
| 2 | 2.09 | 168 | 126 | 84 | 42 | 0.41 |
| 3 | 2.03 | 163 | 122 | 81 | 41 | 0.40 |
| 4 | 1.76 | 141 | 106 | 70 | 35 | 0.35 |
| 5 | 2.38 | 190 | 143 | 95 | 48 | 0.47 |
| 6 | 2.49 | 199 | 149 | 100 | 50 | 0.49 |
| 7 | 2.19 | 175 | 132 | 88 | 44 | 0.43 |
| 8 | 2.23 | 179 | 134 | 89 | 45 | 0.44 |
| 9 | 1.61 | 129 | 97 | 64 | 32 | 0.32 |
| 10 | 0.54 | 43 | 32 | 22 | 11 | 0.11 |
| 11 | 2.66 | 213 | 160 | 106 | 53 | 0.52 |
| 12 | 7.73 | 619 | 464 | 309 | 155 | 1.52 |
| 13 | 2.96 | 237 | 178 | 119 | 59 | 0.58 |
| 14 | 1.87 | 150 | 112 | 75 | 37 | 0.37 |
| 15 | 0.09 | 7 | 6 | 4 | 2 | 0.02 |
| 16 | 1.65 | 132 | 99 | 66 | 33 | 0.32 |
| 17 | 2.67 | 214 | 160 | 107 | 53 | 0.52 |
| 18 | 1.51 | 121 | 91 | 60 | 30 | 0.30 |
| 19 | 1.96 | 157 | 118 | 78 | 39 | 0.38 |
| 20 | 2.28 | 183 | 137 | 91 | 46 | 0.45 |
| 21 | 3.11 | 249 | 187 | 124 | 62 | 0.61 |
| 22 | 5.11 | 409 | 307 | 205 | 102 | 1.00 |
| 23 | 4.39 | 351 | 263 | 176 | 88 | 0.86 |
| 24 | 1.44 | 115 | 86 | 57 | 29 | 0.28 |
| 25 | 15.11 | 1209 | 907 | 605 | 302 | 2.97 |
| 26 | 1.75 | 140 | 105 | 70 | 35 | 0.34 |
| 27 | 2.41 | 193 | 145 | 97 | 48 | 0.47 |
| 28 | 2.70 | 216 | 162 | 108 | 54 | 0.53 |

** **Conversion factors (K. Saito et al., 1990; UNSCEAR 1993 Report) in $nGy\ h^{-1}$ per $Bq\ kg^{-1}$: 0.623 for $^{232}Th$ series, 0.461 for $^{238}U$ series and 0.0414 for $^{40}K$.**

## FIGURE CAPTIONS

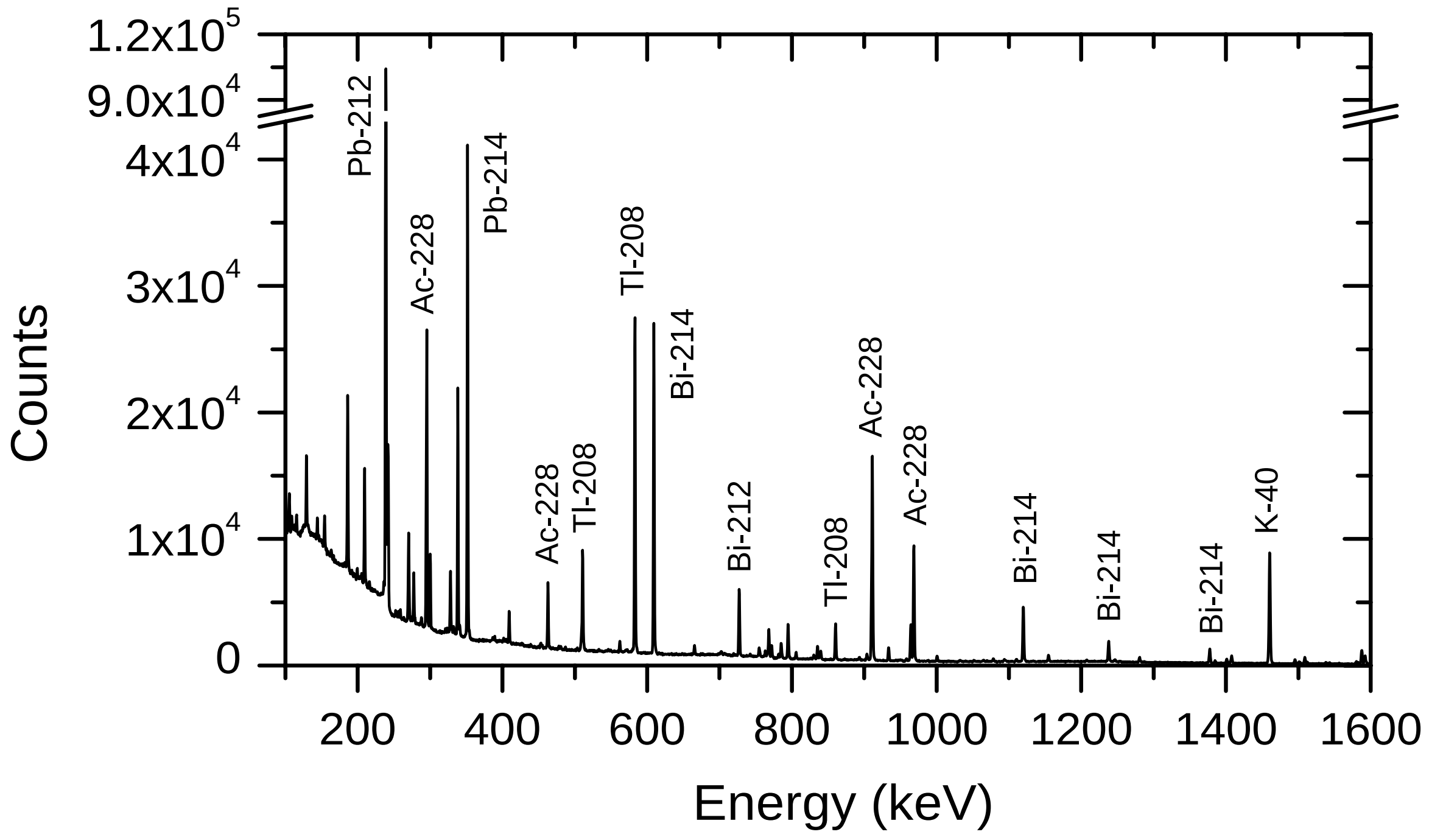

**Figure 1**. Gamma-ray spectrum of a natural tiling rock (“granite”), sample No.11 (“Baltic Brown”). The accumulation time was 37,000 *s*.

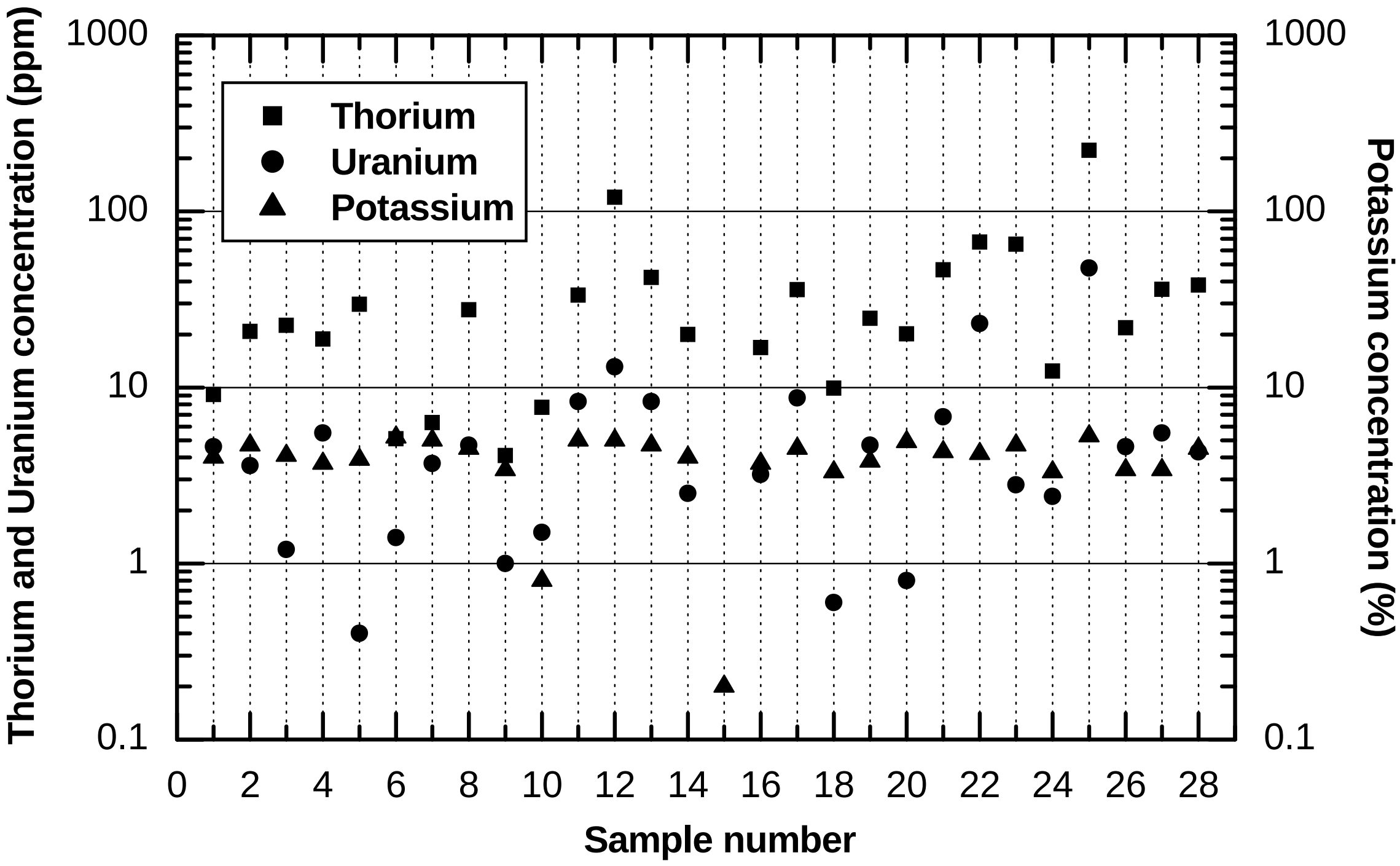

**Figure 2**. Th, U and K elemental concentrations in the samples of "granite" rocks (Table 1) imported and used in Cyprus.

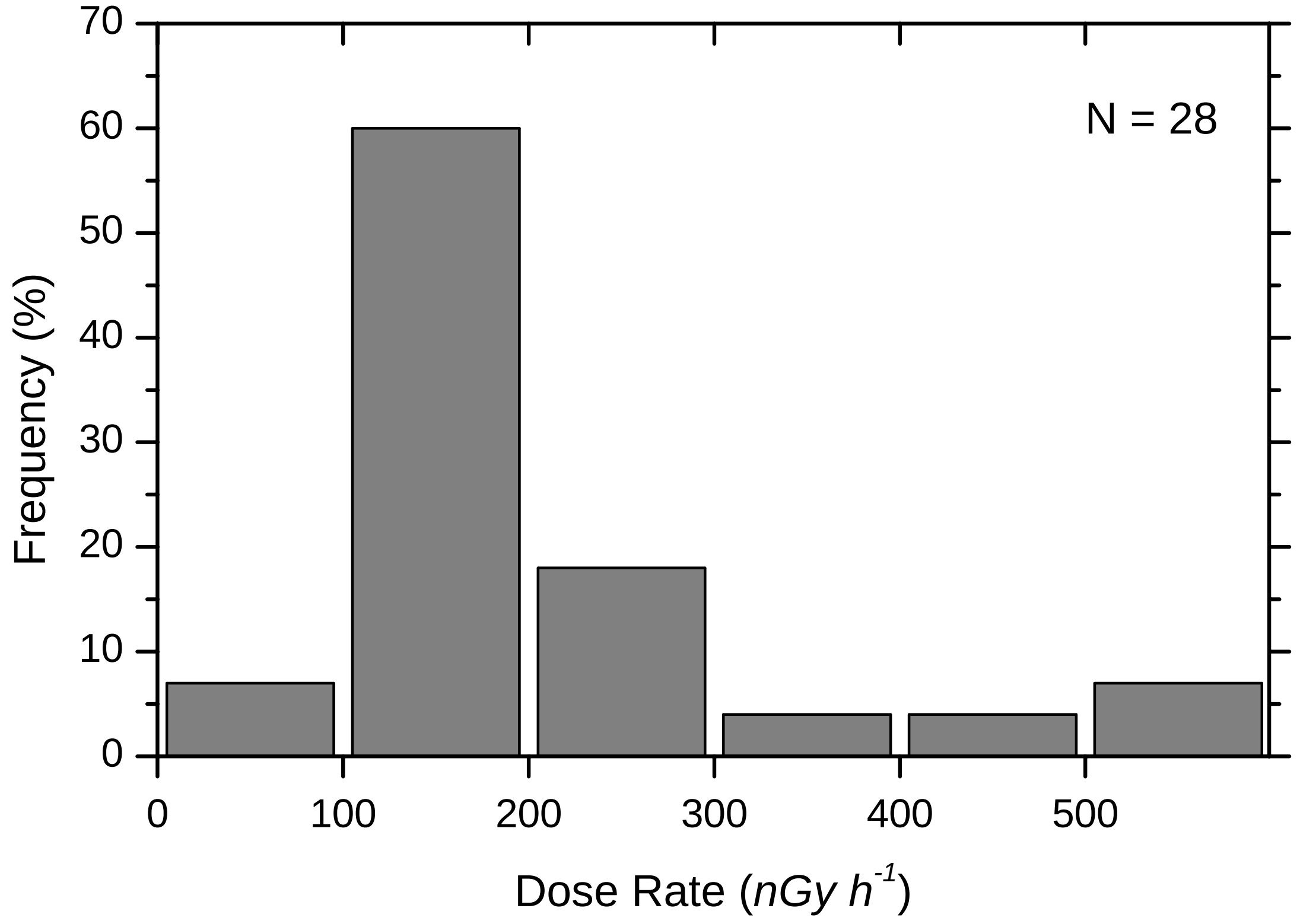

**Figure 3.** Frequency distribution of the total absorbed dose rates for full utilization of the 28 measured “granite” samples. Last column represents those samples, which exhibit dose rates between 500 and 1300 *nGy* $h^{-1}$.

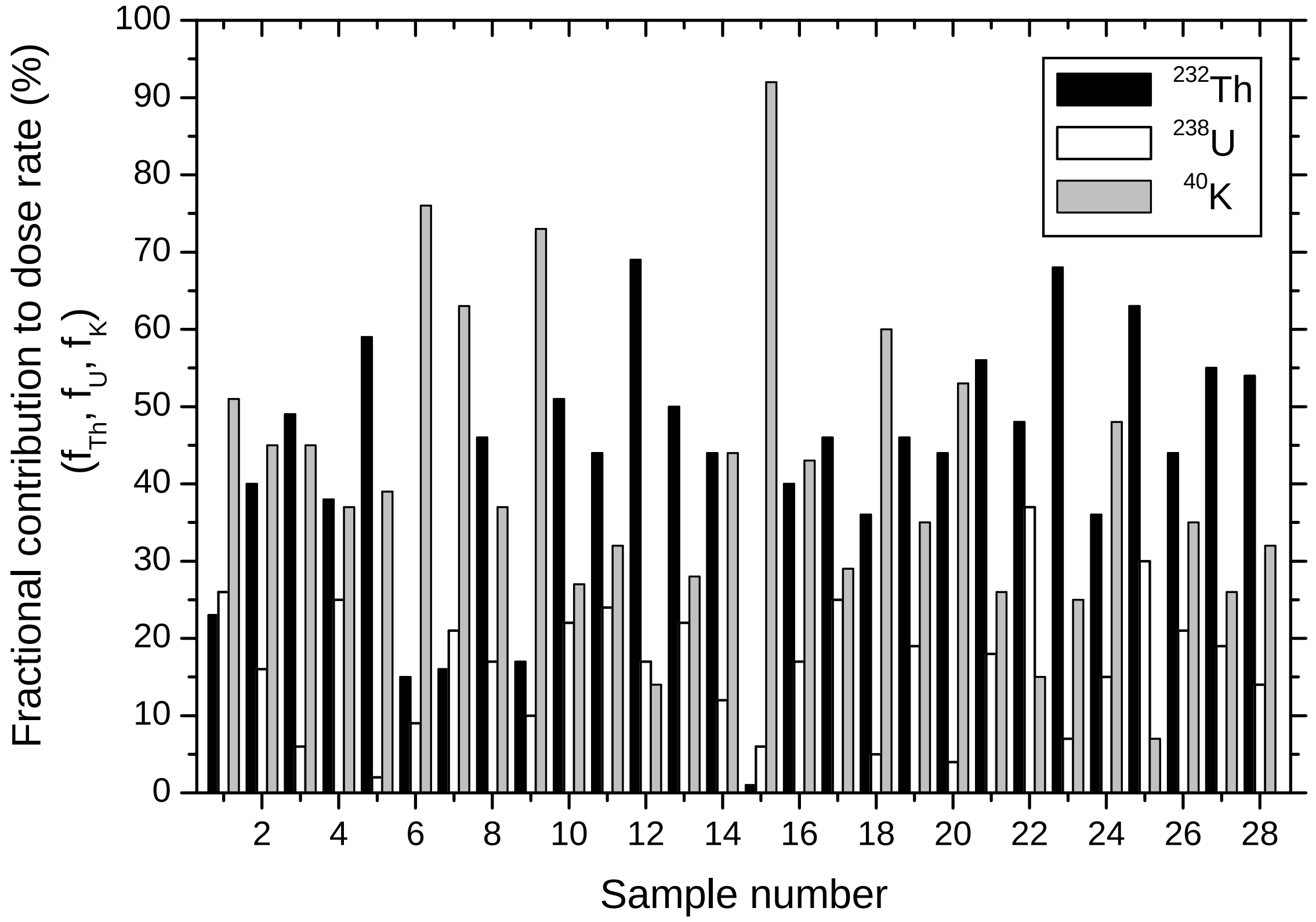

**Figure 4.** Thorium ($^{232}$Th) and uranium ($^{238}$U) decay series and potassium ($^{40}$K) percentage contribution to total absorbed dose rate in air for the measured "granite" samples (Table 1).